\documentclass[%
 reprint,
superscriptaddress,
noeprint,
amsmath,amssymb,
prapplied,
]{revtex4-1}
\usepackage{graphicx}
\usepackage{dcolumn}
\usepackage{bm}
\usepackage{hyperref}
\usepackage{natbib}
\usepackage{subfigure}
\usepackage{tikz}
\usepackage{pgfplots}
\usetikzlibrary{decorations.markings}
\pgfplotsset{width=10cm,compat=1.9}
\pgfplotsset{yticklabel style={text width=2em,align=right}}

\begin{document}


\title{Surface phonon induced rotational dissipation for nanoscale solid-state gears 
}

\author{H.-H. Lin}
 \email{hhlin@nano.tu-dresden.de}
\affiliation{%
 Institute for Materials Science and Max Bergmann Center of Biomaterials, TU Dresden, 01069 Dresden, Germany\\
}
\author{A. Croy}%
\affiliation{%
 Institute for Materials Science and Max Bergmann Center of Biomaterials, TU Dresden, 01069 Dresden, Germany\\
}
\author{R. Gutierrez}
\affiliation{%
 Institute for Materials Science and Max Bergmann Center of Biomaterials, TU Dresden, 01069 Dresden, Germany\\
}
\author{G. Cuniberti}
\homepage{https://nano.tu-dresden.de/}
\affiliation{%
 Institute for Materials Science and Max Bergmann Center of Biomaterials, TU Dresden, 01069 Dresden, Germany\\
}
\affiliation{%
Dresden Center for Computational Materials Science, TU Dresden, 01062 Dresden, Germany\\
}
\affiliation{%
Center for Advancing Electronics Dresden, TU Dresden, 01062 Dresden, Germany\\
}

\date{\today}

\begin{abstract}
Compared to nanoscale friction of translational motion, the mechanisms of rotational friction have received less attention. Such motion becomes an important issue for the miniaturization of mechanical machineries which often involve rotating gears. In this study, molecular dynamics simulations are performed to explore rotational friction for solid-state gears rotating on top of different substrates. In each case, viscous damping of the rotational motion is observed and found to be induced by the pure van-der-Waals interaction between gear and substrate. The influence of different gear sizes and various substrate materials is investigated. Furthermore, the rigidities of the gear and the substrate are found to give rise to different dissipation channels. Finally, it is shown that the dominant contribution to the dissipation is related to the excitation of low-frequency surface-phonons in the substrate.
\end{abstract}

\maketitle


\section{\label{sec:Introduction}Introduction}
Friction on the nanoscale is a topical subject of great interest from both a practical as well as a fundamental point of view. However, extrapolating well-known laws from the macroscopic domain to nanoscale friction phenomena is not straightforward and often, the observations in this regime are in contrast to macroscopic behavior. Thus, it was found that Armonton's law and Coulomb's law cannot be directly applied to nanoscale frictional processes \cite{Sankar2020}. Consequently, in the context of nanotribology, sliding friction has been extensively studied from different perspectives \cite{Vanossi2013,Gnecco2006,Perssson1995,Wen2017,Manini2016,Sankar2020}, including numerical studies of different phenomenological models \cite{Tomlinson1929,Borner,Panizon2018} as well as various experimental techniques \cite{Lodge2016,Smith1996,Walker2012,Binnig1986,Binnig1982,Binnig1987,Balakrishna2014}.

Prototypical investigations involve nanoparticles moving on a surface and interacting with it via van-der-Waals forces \cite{Guerra2010}. The friction resulting from the mutual interaction is typically discussed in the context of translational motion. Here, different friction regimes have been identified depending on the center-of-mass velocity. In the low-speed regime ($v\ll $ 1000 cm/s) \cite{Tomassone1997a}, the system is close to equilibrium and friction is governed by Brownian motion, in which the stochastic behavior is dominant. In the high-speed regime \cite{Tomassone1997a,Guerra2010}, the system is strongly out of equilibrium and the friction force is proportional to the number of collisions with surface corrugations per unit time \cite{Guerra2010}, and thus the friction becomes viscous.

Compared to friction in translational motion, the case of purely rotational friction has received less attention. Since high tangential velocity (outer) and low tangential velocity (inner) atoms are present at the same time and both sets are forced to move concurrently, it is not obvious to what extent the results for translational friction can be applied in a purely rotational setting.
In this context, rotating gears are of particular interest for the miniaturization of mechanical devices. They are key components of many potential applications, such as the implementation of nanoscale analog computing devices (analogous to the Pascaline \cite{Roegel2015}) or in nanorobotics \cite{Weir2005}. 

Over the last years, gears have been successfully downsized to the nanoscale. The miniaturization was achieved through bottom-up approaches by synthesizing molecular gears \cite{Gisbert2019}, and by top-down methods, such as focused ion beam \cite{JuYun2007} or electron beam \cite{Deng2011,Yang2014} techniques, allowing it to etch solid-state gears. For molecule gears, many issues, such as motor effects \cite{Lin2019,PES,Croy2012,Stolz2020,Pshenichnyuk2011}, manipulation \cite{Eisenhut2018,Manzano2009,Light1999,Zhang2019} and transmission \cite{Lin2019a,WeiHyo2019,Lin2020}, have been studied. For solid-state gears, a recent study has focused on transmission of rotational motion in gear trains \cite{Lin2020}.

\begin{figure}[t]
\centering
 \includegraphics[width=0.45\textwidth]{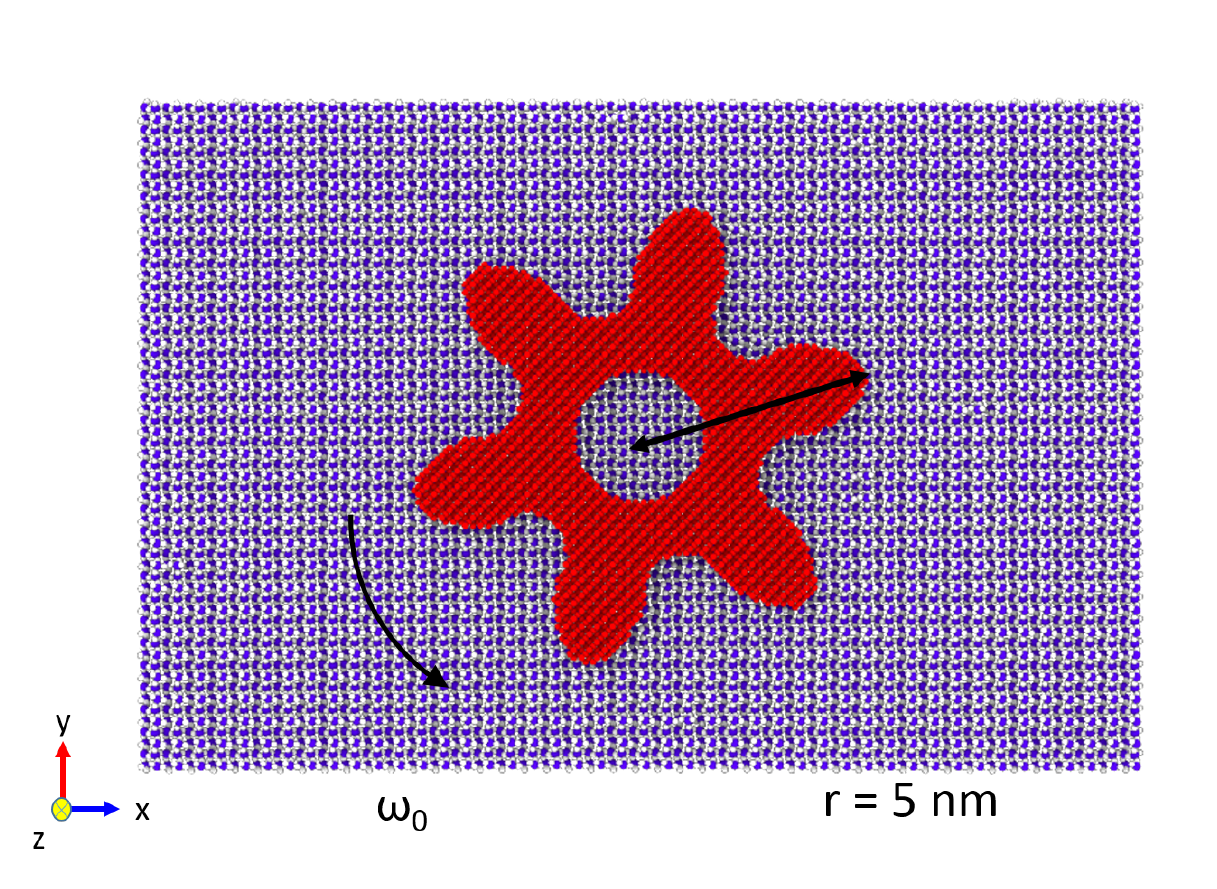}
 \caption{Schematic illustration of a diamond-based solid state gear with radius $r=5$ nm rotating with angular velocity $\omega_0$ on top of a $\alpha$-cristobalite SiO$_2$ (001) substrate.}
 \label{Fig:scheme}
\end{figure}
In this work, we focus on diamond-based solid-state gears rotating on a substrate (see Fig.\ \ref{Fig:scheme}). We use various combinations of gears and substrate materials to address the associated questions about the dependence of rotational friction on the crystal structure of the substrate and the gear size. For this type of problem, it is suitable to use classical molecular dynamics (MD) simulations as in previous studies of translational friction \cite{Tomassone1997a,Guerra2010}. This approach allows us to take a large number of atoms into account and, moreover, to reach relevant time scales \cite{Persson1996} in the range of $100$ps up to $1$ns.

The paper is organized as follows: In Sec.\ \ref{sec:Formalism}, we introduce the setup of the gear-substrate system and also the technical details of the MD simulations. In Sec.\ \ref{sec:Results}, we study a diamond-based solid-state gear on a SiO$_2$ substrate with given initial angular velocity and investigate its slowing down. This is followed by a study of dissipation with different combinations of gear sizes and 
substrate materials. Then, we investigate various setups involving different degrees of conformational freedom of the gear plus substrate system to explore its influence on friction. Furthermore, we study the effect of the thickness of the substrate and the resulting phonon excitation by using kinetic energy spectra. It shows that the low frequency surface phonon has been excited during the gear rotation.

\section{\label{sec:Formalism} Model system and computational approach }
In this section, we introduce the system setup and provide details on the MD simulations.

\subsection{\label{sec:setup}Setup}
\begin{figure}[t!]
\centering
 \includegraphics[width=0.5\textwidth]{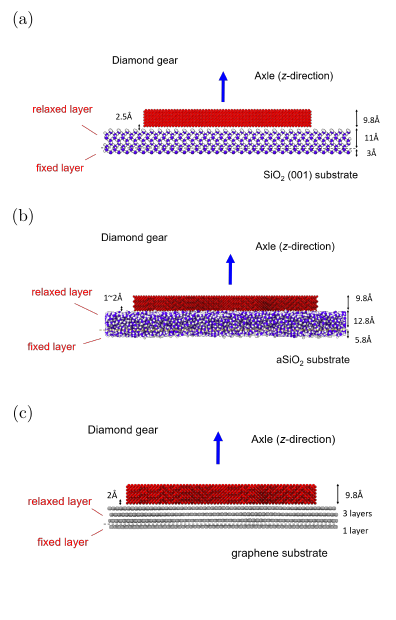}
 \caption{Scheme of the $5$ nm diamond-based solid-state gear on top of the (a) $\alpha$-cristobalite SiO$_2$ with size 23.0 nm$\times$15.3 nm$\times$1.39 nm and separation distance 2.5\AA;(b) amorphous SiO$_2$ with dimension 28.2 nm$\times$16.9 nm$\times$1.86 nm and separation distance around $1$ to 2\AA{ }and (c) 4-layer graphene with size 21.8 nm$\times$12.6 nm$\times$ 4 monolayers and separation distance 2\AA.}
 \label{Fig:substrate}
\end{figure}

To start our study of how energy can be dissipated from the gear into the surface, we consider the system shown in Fig.\ \ref{Fig:scheme}. Here, we put a diamond-based solid state gear with radius $r=5$ nm and thickness $9.8$\AA{} (with $8544$ carbon atoms) on top of a SiO$_2$ (001) surface (see Fig.\ \ref{Fig:substrate} (a)). The geometry of the solid state gear was created on the basis of the general algorithm for involute \footnote{The involute shape was proposed by Euler as an optimized geometry for classical rigid-body transmission.} spur gears \cite{Norton2010}, which can be implemented by using the Open Visualization Tool \cite{Stukowski2010} (OVITO) to cut the gear from a bulk diamond crystal. For the substrate, we create an $\alpha$-cristobalite SiO$_2$ (001) surface (with size 23.0 nm$\times$15.3 nm$\times$1.39 nm, which contains 32400 atoms) from Visual Molecular Dynamics \cite{Humphrey1996} (VMD) with periodic boundary conditions in $x$ and $y$ directions. We then combine the gear and substrate together with an initial vertical separation of 2.5\AA{}. To prevent lateral and vertical motion, we fix the bottom layer of the substrate (with thickness $3$\AA{}) and let the remaining part (a region with thickness 11\AA{}) to relax. Then, a geometry optimization is performed using the conjugate gradient method within the MD simulation software, which will be discussed in detail later.

To compare different materials, we prepare the system with the same diamond gear but with two different substrate materials: amorphous SiO$_2$ and graphene (see Figs.\ \ref{Fig:substrate} (b) and (c)). For the amorphous SiO$_2$, we create a substrate with dimension 28.2 nm$\times$16.9 nm$\times$1.86 nm and set the initial vertical separation to be approximately $1$ to $2$ \AA{} (since the surface has many asperity points, the distance is not well-defined). To prevent net translation of the substrate, we fix the bottom layer with a thickness 5.8\AA{} to mimic the semi-infinite bulk material. Then, we let only the remaining layer with thickness of 12.8\AA{} layer to be relaxed. For graphene, the substrate size is 21.8 nm$\times$12.6 nm$\times$ including 4 monolayers with initial vertical separation of 2\AA{}. The bottom monolayer is kept fixed and the other 3 layers are allowed to relax.
\subsection{\label{sec:MD}Molecular dynamics}
Once the system is setup, the next step is to define the simulation protocol for the MD. Here, we use the Large-scale Atomic/Molecular Massively Parallel Simulator (LAMMPS) \cite{Plimpton1995} for implementing the MD simulations. 

For the force fields, we choose the Tersoff potential \cite{Tersoff1988} inside the SiO$_2$ substrate. Inside diamond, we use the adaptive intermolecular reactive empirical bond order (AIREBO) potential \cite{Stuart2013}, which is an established bond-order potential specifically developed to describe allotropes of carbon. Between the gear and substrate, we specify a 12-6 type Lennard-Jones potential $V_{\text{LJ}}(r)$:
\begin{equation}
V_{\text{LJ}}(r)=4\epsilon\left[\left(\frac{\sigma}{r}\right)^{12}-\left(\frac{\sigma}{r}\right)^6\right]\;,
\end{equation}
where $r$ is the interatomic distance. Here we choose the LJ parameters \cite{Inui2017}: $\epsilon_{\text{C-Si}}=8.91$ meV, $\epsilon_{\text{C-O}}=3.44$ meV and $\sigma_{\text{C-Si}}=3.629$ \AA{}, $\sigma_{\text{C-O}}=3.275$ \AA{}. Note that between the diamond gear and the graphene substrate, the van-der-Waals interaction is automatically included in the AIREBO potential among carbon atoms \cite{Stuart2013}. To fix the rotation axle, we attach a spring with high spring constant $k=1.6\cdot 10^4$ N/m (1000 eV/\AA$^2$) to the gear center of mass and we fix the temperature $T=10$K with the substrate subject to the canonical ensemble implemented by the Nosé–Hoover thermostat \cite{Nose1984,Hoover1985}.

\section{\label{sec:Results}Results}
In this section, we investigate the friction process by considering the relaxation of the angular velocity via pure Lennard-Jones interactions between gear and substrate. We then address the issue of how different combinations of gear sizes and substrate materials affect the friction. To investigate the dissipation channels in the system, we restrict the dynamics to rigid-body motion for the gear and the substrate for several different cases: rigid gear with deformable substrate, rigid substrate with deformable gear, and both rigid. Finally, we study the influence of the thickness of the substrate relaxed layer and compute the corresponding kinetic energy spectra.

\begin{figure}[t]\centering
 \includegraphics[width=0.45\textwidth]{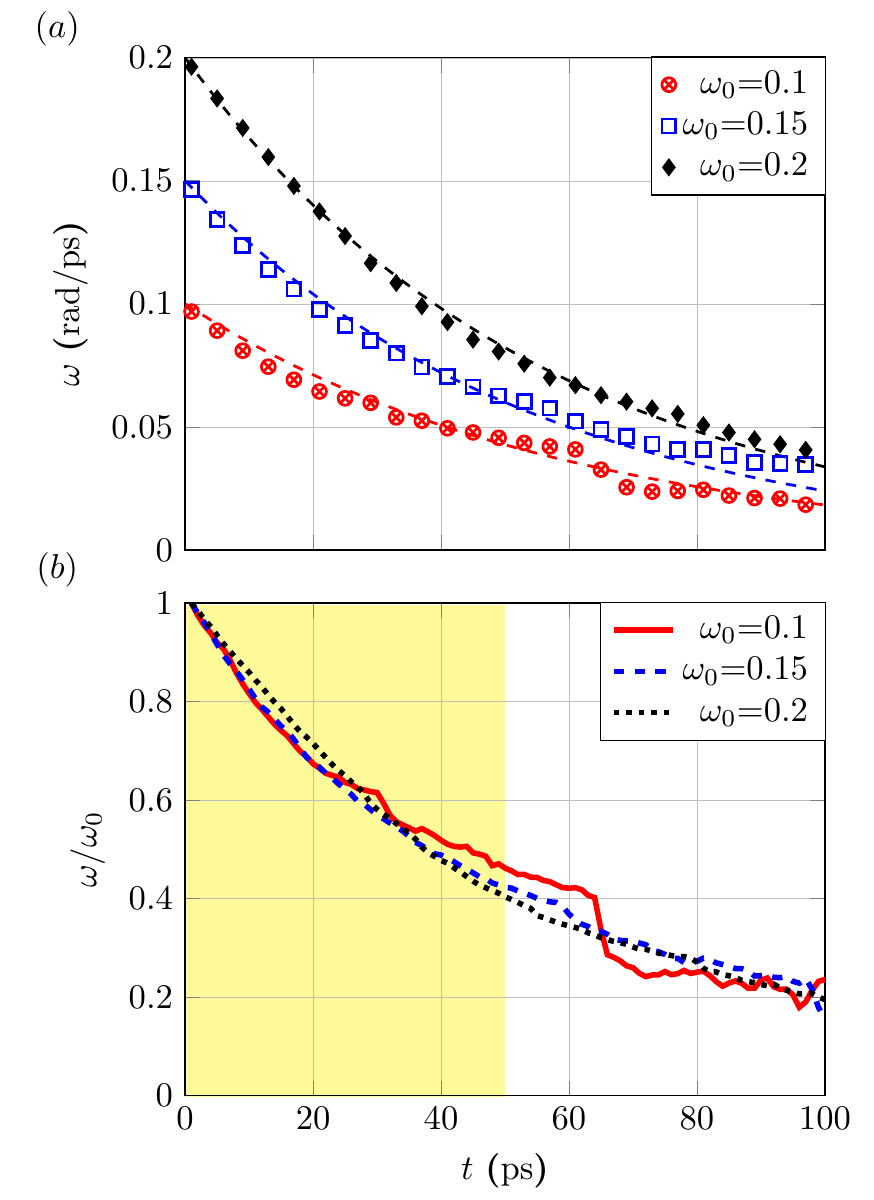}
 \caption{(a) Angular velocity (exponential fits indicated by dashed lines) and (b) normalized angular velocity of the $5$ nm diamond gear on top of $\alpha$-cristobalite SiO$_2$ undergoes a relaxation within the simulation time $t=100$ ps with different initial angular velocity $\omega=0.1$, $0.15$ and $0.2$ rad/ps, respectively. The highlighted yellow region (from $0$ to $50$ ps) is used for extracting the velocity relaxation time $\tau$ by fitting to an exponential function $\propto e^{-t/\tau}$.}
 \label{Fig:viscous}
\end{figure}

\subsection{Viscous dissipation and rotational motion}
Considering that the atoms inside a rotating gear are moving with higher tangential velocity in average, we may expect that the rotational friction still corresponds to the high tangential velocity friction regime as found for translational motion \cite{Guerra2010}. Accordingly, a gear with instantaneous angular velocity $\omega$ is subject to a non-conservative torque equal to $-\gamma \omega$: 
\begin{equation}
 I\dot{\omega}=-\gamma \omega\;,
\end{equation}
where $\gamma$ is the damping coefficient and $I$ is the gear moment of inertia. Therefore, in a first step, we give an initial angular velocity to the gear (which rotates with respect to a fixed center of mass) and check if this viscous dissipation law will emerge as a result of van-der-Waals interactions between the gear and the substrate at fixed temperature. To be specific, we set several different initial angular velocities $\omega_0 = 0.1$, $0.15$ and $0.2$ rad/ps, and then let the system relax during $100$ ps. In Fig.\ \ref{Fig:viscous}(a), the velocity relaxation of a rotating gear due to the interaction with the substrate is shown. One can see that the frictional behavior agrees with viscous damping rather well, which can be further seen from the plot of the normalized angular velocity in Fig.\ \ref{Fig:viscous}(b). Additionally, the angular velocity relaxation for the gear weakly depends on the initial velocity, except for a small fluctuation in the case of $\omega_0=0.1$ rad/ps (there is a kink around $65$ ps due to the internal deformation). We can extract the corresponding relaxation time $\tau$ using the region from $0$ to $50$ ps (highlighted in yellow), by fitting the curve to the exponential function $\propto e^{-t/\tau}$ (dashed lines in Fig.\ \ref{Fig:viscous}(a)), which gives the relaxation times $58.62$, $54.55$ and $56.22$ ps for $\omega_0=0.1$, $0.15$ and $0.2$ rad/ps, respectively. Note that the relaxation time is related to the damping coefficient $\gamma$ through the relation $\tau=I/\gamma$. Another observation is that the gear with size $r=5$ nm behaves like a 
deterministic rotor instead of a Brownian rotor, since the stochastic behavior is suppressed with increasing gear size. This implies that the friction of the $5$nm involute shape gear is in the high average tangential velocity dissipation regime. 

\subsection{Effect of substrate material and gear size}
Apart from the gear on top of the crystalline SiO$_2$ substrate, one may ask how different combinations of gear and substrate would affect the friction. To avoid too many variables, we use diamond gears and gradually reduce their size. For the substrate, we change the material from $\alpha$-cristobalite SiO$_2$ to amorphous SiO$_2$ and finally to few-layers graphene, as shown in Fig.\ \ref{Fig:substrate}.
\begin{figure}[h]
 \includegraphics[width=0.45\textwidth]{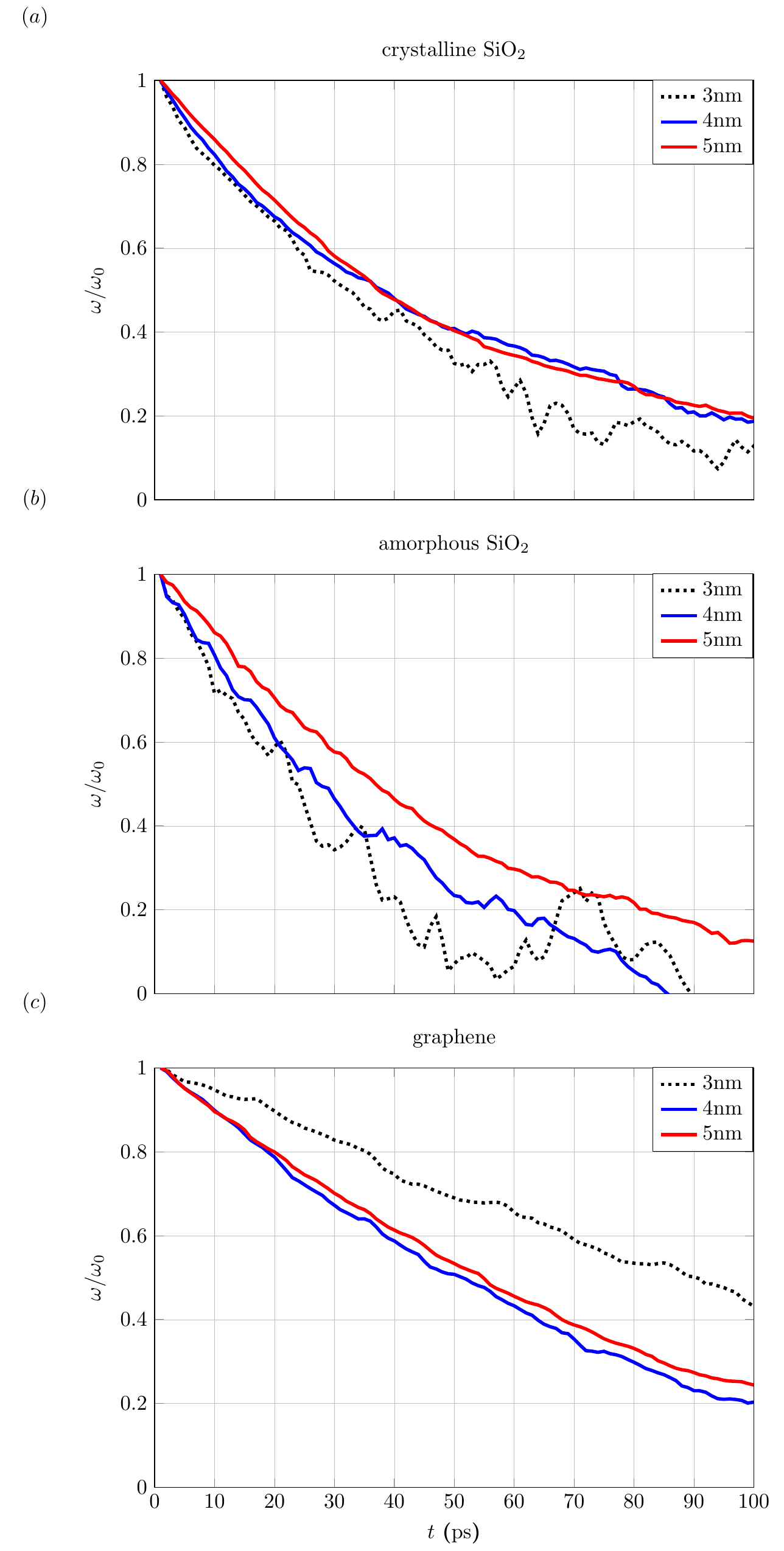}
 \caption{The normalized angular velocity of the diamond-based solid-state gear with initial angular velocity $\omega_0=0.2$ rad/s for different gear radius $r=3,4$ and 5 nm on top of (a) crystalline SiO$_2$, (b) amorphous SiO$_2$ and (c) 4-layer graphene, respectively.}
 \label{Fig:size_material}
\end{figure}
\begin{table}[t]
\begin{ruledtabular}
\begin{tabular}{cccc}
\textrm{Size }&
\textrm{3nm}&
\textrm{4nm}&
\textrm{5nm}\\
\colrule
$I$ & 1.98 & 6.22 & 19.7\\
\colrule
c-SiO$_2$ & 46.87 (4.22)& 53.09 (11.72)& 56.22 (34.86)\\
a-SiO$_2$ & 29.92 (6.62) & 39.25 (15.85) & 54.20 (36.16)\\
graphene & 135.49 (1.46)& 69.76 (8.92)& 76.22 (25.71)\\
\end{tabular}
\end{ruledtabular}
\caption{\label{tab:table1}%
Moments of inertia $I$ (in units of 10$^{-40}$ kg$\cdot$m$^2$) and angular velocity relaxation times in units of picosecond for different combinations of gear size and substrate material. The corresponding damping coefficients $\gamma$ in units of 10$^{-30}$ kg/ps are given in parentheses.
}
\end{table}

With the above setup, we perform a $100$ ps MD simulation at temperature $T=10$ K with the initial angular velocity $\omega_0=0.2$ rad/ps. The results are shown in Fig.\ \ref{Fig:size_material}. We change the gear radius from $5$nm to $3$nm for crystalline SiO$_2$ ($\alpha$-cristobalite), amorphous SiO$_2$ and graphene in Fig.\ \ref{Fig:size_material} (a), (b) and (c), respectively. One can see that as the gear size reduces to $3$nm (black dotted line), the gear motion becomes fluctuating, especially for amorphous SiO$_2$ and graphene. This is related to the fact, that when the rotational kinetic energy becomes smaller, the gear will be more sensitive to surface vibrations, which are determined by $k_BT$, and, therefore, the gear starts entering the low average tangential velocity dissipation regime, in which Brownian behavior becomes dominant. Next, we extract the corresponding relaxation times from the curves and summarize them in Table \ref{tab:table1}. One can clearly see that for crystalline SiO$_2$, the relaxation times are approximately independent of the gear size. However, the relaxation time for amorphous SiO$_2$ shows that as the gear size reduces the relaxation becomes much shorter. For graphene, it shows a rather surprising behavior. For $4$nm and $5$nm, the relaxation time is around $70$ ps, but for the $3$ nm gear it suddenly increases to $135$ ps, which could indicate a super-lubricant behavior in this case. Compared to SiO$_2$, the graphene substrate is essentially flat on the length scale given by the smallest gear. Thus, the coupling between the gear and the substrate is less effective in this case.  Based on our simulations, amorphous SiO$_2$  shows the largest friction coefficient, crystalline SiO$_2$ has an intermediate one, and graphene shows the smallest friction coefficient. 

\subsection{Dissipation and available degrees of freedom}

Although dissipation is directly related to the interaction with the substrate, the detailed mechanism for the energy transfer is still unclear. Therefore, in this section, we discuss how the different number of available degrees of freedom can affect the dissipation. Here, we will discuss four different cases: case 1, gear and substrate are both rigid bodies; case 2, deformable gear and rigid substrate; case 3, rigid gear and deformable substrate and case 4, gear and substrate are deformable. Case 1 is trivial as long as the gear motion is still a planar rotation, since the rotating gear will move in a potential energy surface, which is defined collectively by the atoms in the frozen substrate. Therefore, the gear will keep rotating or oscillating permanently without any dissipation. In case 2, the gear is allowed to deform but the substrate is frozen. Therefore, the potential energy surface is still the same as in case 1. However, the rotating gear can now also transfer energy to the internal degrees of freedom of the gear, leading to a deformation of the gear. For the case 3, the gear is rigid but the substrate is deformable. Therefore, the potential energy surface is no longer well-defined and the force acting on the gear is not conservative, since the energy can be transported away from the surface by bulk phonons. However, if the angular velocity of the gear is sufficiently high, it will experience a force resulting from the averaged motion of the surface. Finally, case 4 is a mixture of cases 2 and 3, which has two dissipation channels at the same time. 
\begin{figure}[t]\centering
 \includegraphics[width=0.45\textwidth]{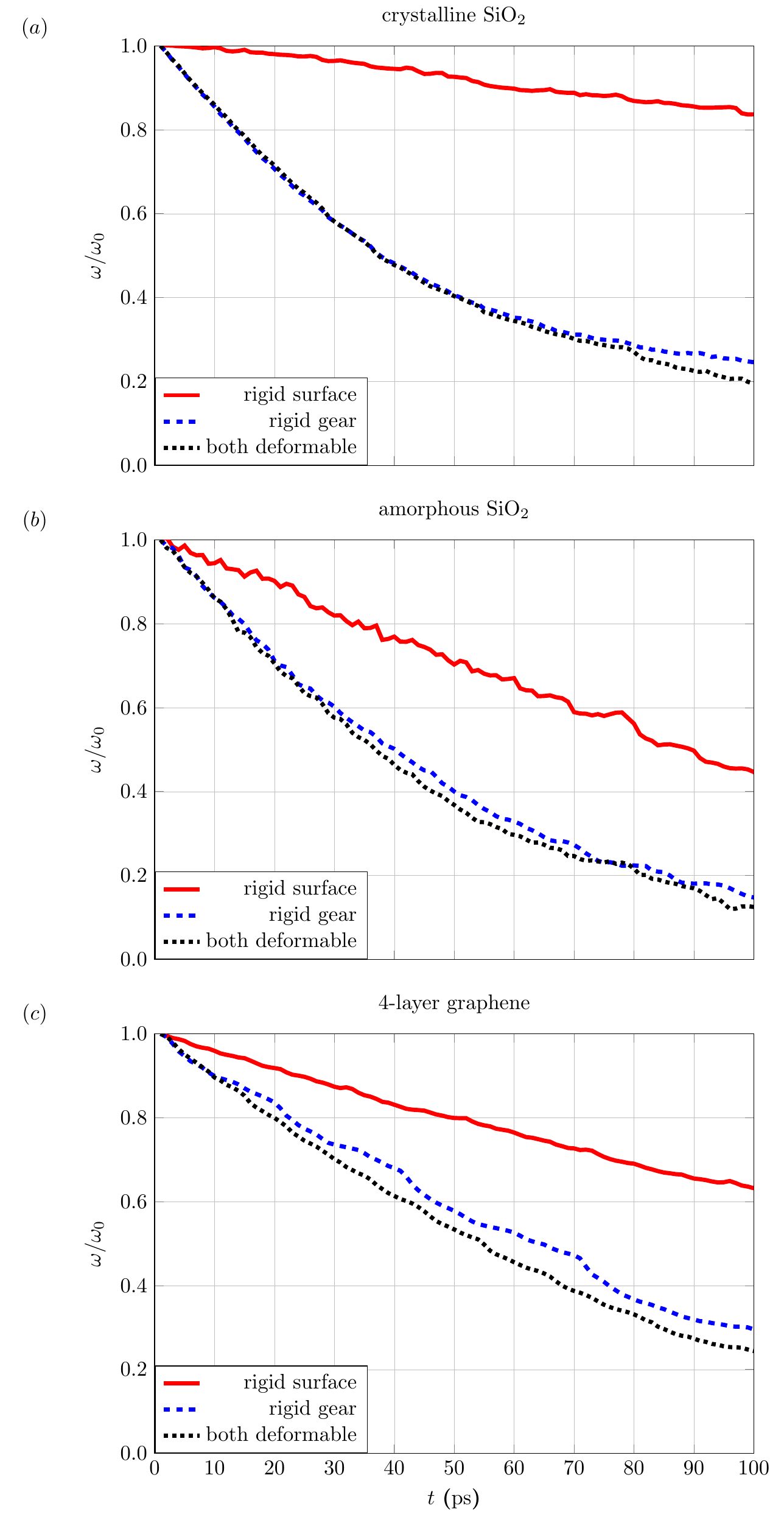}
 \caption{The normalized angular velocity with initial angular velocity $\omega=0.2$ rad/ps for different cases of available degrees of freedom: rigid surface, rigid gear and both deformable for (a) crystalline SiO$_2$, (b) amorphous SiO$_2$ and (c) graphene.}
 \label{Fig:DOF}
\end{figure}
The results of cases 2 to 4 are shown in Fig.\ \ref{Fig:DOF}. We compare the angular velocity relaxation among different cases and materials with the same initial angular velocity $\omega=0.2$ rad/ps, gear radius $r=5$ nm and temperature $T=10$ K (if the substrate is deformable) within $100$ ps. One can see that the angular velocity relaxation corresponding to the deformable gear and substrate is faster than that of the rigid gear or rigid substrate, which provides a strong evidence that the whole dissipation channel is composed of channel 1 and channel 2, which is independent of substrate material (except for the small angular velocity regime of amorphous SiO$_2$ substrate due to the location of asperity points). Also, the channel 2 dissipates more efficiently than channel 1, which implies that surface vibrations are more important than gear deformation for the substrates considered in this study.

\subsection{Thickness dependence and surface phonon excitation}
\begin{figure}[t]\centering
 \includegraphics[width=0.45\textwidth]{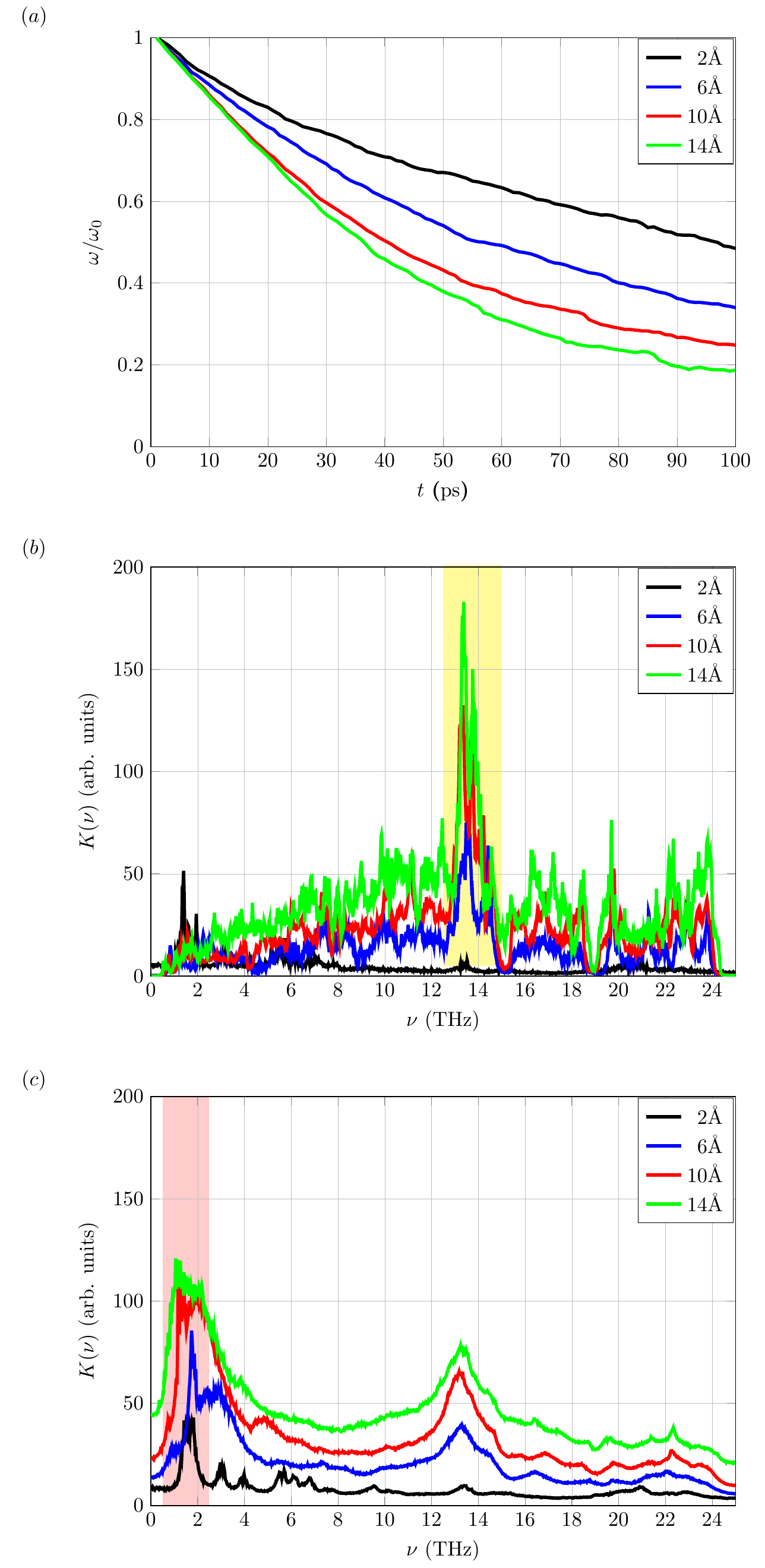}
 \caption{(a) Relaxed layer thickness dependence of angular velocity relaxation. Kinetic energy spectra of substrate (b) without and (c) with gear rotation within $50$ ps.}
 \label{Fig:thickness_phonon}
\end{figure}

Since we have found that the dissipation is mainly coming from surface vibrations, it is interesting to see how the dissipation can gradually emerge when the thickness of the relaxed layer in Fig.\ \ref{Fig:substrate} is changed. Note that the overall thickness of the substrate in the simulations is always the same, only the thickness of the fixed layer is changed. The results for crystalline SiO$_2$ are shown in Fig.\ \ref{Fig:thickness_phonon} (a). One can see that as the thickness $d$ of the relaxed layer increases, the angular velocity relaxation time $\tau$ becomes shorter. However, the thickness dependence is only important when $d$ is not too large. As one can see when $d$ approaches 14\AA{}, the decay rate starts to saturate. 

To get additional insight, we define the kinetic energy spectrum $K(\nu)$ \cite{Meshkov1997,Lee1993} given by
\begin{equation}
 K(\nu)=\sum_{i\in \text{R}}K_i(\nu)=\sum_{i\in \text{R}}\frac{1}{2}m_i|\mathbf{v}_i(\nu)|^2 \;,
\end{equation}
where $\mathbf{v}_i(\nu)$ is the Fourier transform of the velocity of the $i^{\text{th}}$ atom in the relaxed layer (R) and $\nu$ is the frequency. To identify the features arising purely from rotational dissipation, we show the kinetic energy spectra for the cases without and with gear rotation. 
Using a time-step of $20$ fs and a total simulation time of $50$ ps, gives a maximal frequency equal to $25$ THz and a resolution of $0.02$ THz for the spectra.
The results are shown in Figs.\ \ref{Fig:thickness_phonon} (b) and (c), respectively. In Fig.\ \ref{Fig:thickness_phonon} (b), one can see that as the thickness increases, a feature around $13$ THz appears, This feature can be associated with the van-Hove singularity of the phonon density of states for the crystalline SiO$_2$ thin film. Without gear rotation the system is in a state of thermal equilibrium for which the equipartition theorem applies and one finds \cite{Meshkov1997,Lee1993}:
\begin{equation}
 \rho(\nu)\propto\frac{K(\nu)}{\frac{3}{2}N_Rk_BT}\;,
\end{equation}
where $\rho(\nu)$ is the phonon density of states (PDOS) and $N_R$ is the number of atoms in the relaxed layer. Moreover, $k_B$ denotes the Boltzmann constant and $T$ is the temperature. The peak position at $13$ THz also agrees with the PDOS from density functional perturbation theory (DFPT) calculations \cite{Wehinger2015}.

Now we can compare the kinetic energy spectrum with that during the first $50$ ps shown in the yellow region in Fig.\ \ref{Fig:viscous} (b), which is the duration of the dissipation process. The result is shown in Fig.\ \ref{Fig:thickness_phonon}
(c). One can see that when the gear is rotating and dissipating the original peak at $13$ THz is broadened, which means that the lifetime of this vibrational mode becomes shorter. At the same time, the gear also excites low frequency modes around $1$ to $2$ THz. Moreover, the amplitude of the spectrum increases as the thickness becomes larger, which indicates that more energy has been dissipated into the substrate and phonons are excited. However, the thickness effect becomes less important since the amplitudes do not change too much as the thickness gets larger than $10$ \AA{}. This implies that only surface phonons are excited which is also consistent with our previous finding for the angular velocity relaxation in Fig.\ \ref{Fig:thickness_phonon}
(a). 

\subsection{Phonon mode visualization}
\begin{figure}[t]\centering
 \includegraphics[width=0.45\textwidth]{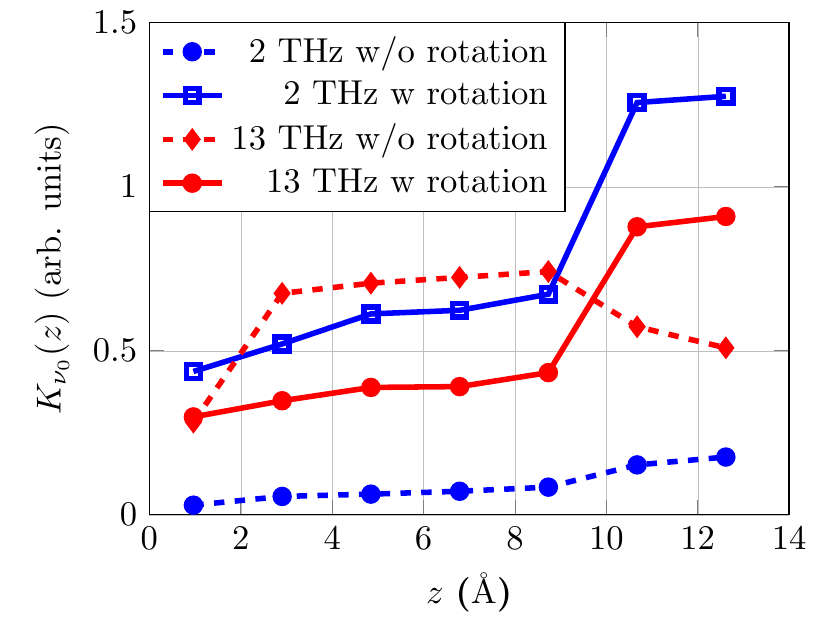}
 \caption{The vertical distribution of the average kinetic energy spectrum $K_{\nu_0}(z)$ of the substrate with the relaxed layer thickness 14\AA{}, where $z$ is the position relative to the bottom of the crystalline SiO$_2$ substrate. The blue and red lines show $K_{\nu_0}(z)$ for $\nu_0=2$ and $\nu_0=13$ THz, respectively; the dashed line and solid line represent $K_{\nu_0}(z)$ before and during the gear rotation.}
 \label{Fig:vertical_distribution}
\end{figure}

To better understand the excited modes  before and during the gear rotation, we filter the kinetic energy spectrum for a given frequency $\nu_0$. In order to do this, we apply a band filter to the kinetic energy spectrum for the $i^{\text{th}}$ atom as follows:
\begin{equation}
 K_{i}(\nu_0)=\int_{-\infty}^{\infty}K_i(\nu) \mathcal{W}(\nu-\nu_0)d\nu\;.
\end{equation}
In practical terms, we use a rectangular window function:
\begin{equation}
 \mathcal{W}(\nu-\nu_0)\approx \frac{\theta[\nu-(\nu_0-\Gamma/2)]-\theta[\nu-(\nu_0+\Gamma/2)]}{\Gamma}\;,
\end{equation}
where $\Gamma$ is the bandwidth of the window around $\nu_0$ and the function $\theta(\nu)$ is the step function. 
\begin{figure*}[t]\centering
 \includegraphics[width=0.95\textwidth]{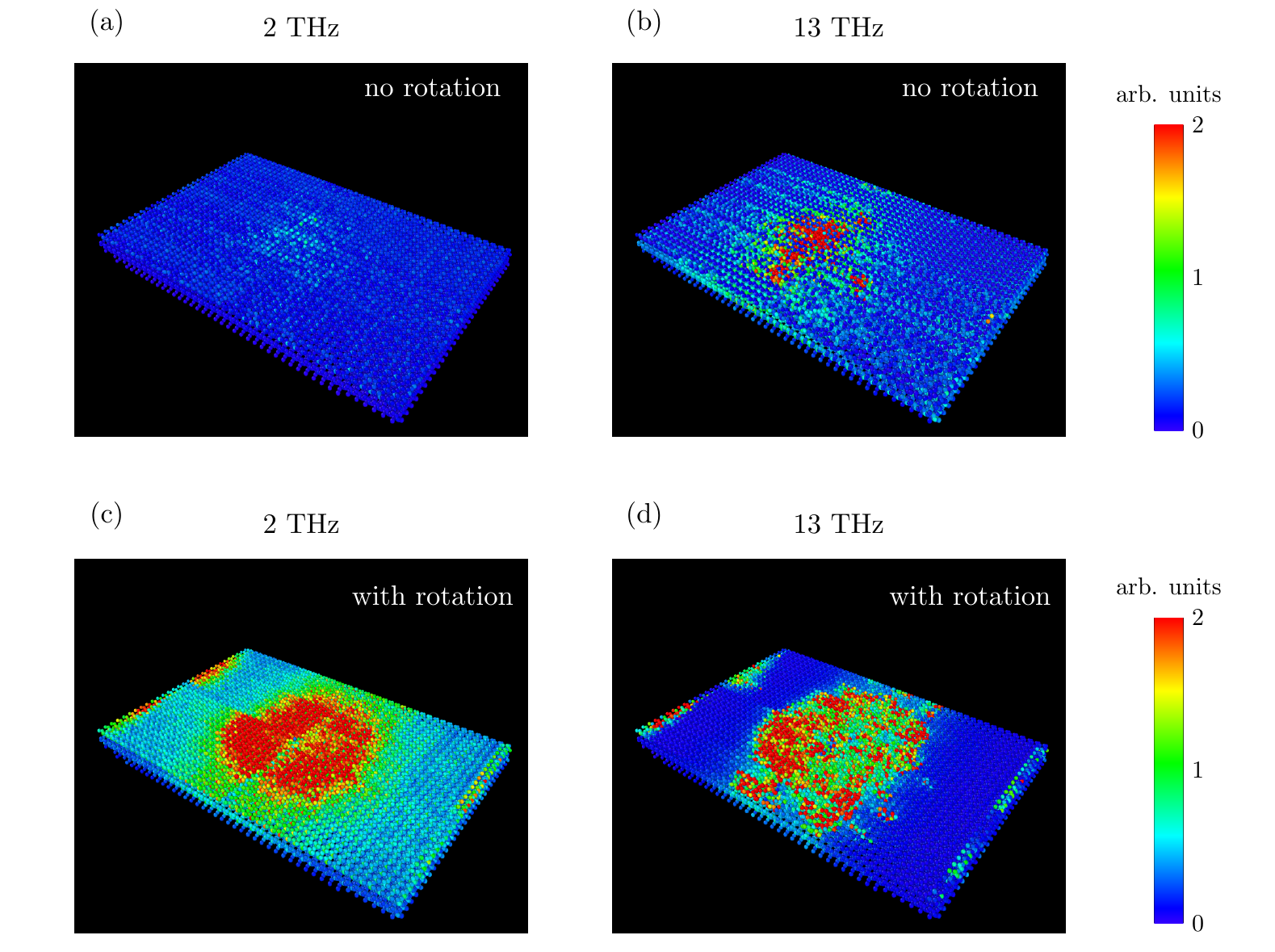}
 \caption{3D topography of the intensity from the kinetic energy spectrum $K_j(\nu_0)$ with contribution around the frequencies (with bandwidth $\Gamma=1$ THz): (a) $\nu=2$ THz without rotation, (b) $\nu=13$ THz without rotation, (c) $\nu=2$ THz during rotation and (d) $\nu=13$ THz during rotation.}
 \label{Fig:phonon_modes}
\end{figure*}
The respective windows are highlighted in Fig.\ \ref{Fig:thickness_phonon} (b) and (c). The features around $2$ and $13$ THz are of particular interest. Therefore, we compare the corresponding modes at these frequencies before and during gear rotation for the crystalline SiO$_2$ substrate with a relaxed layer thickness of
$14$\AA{}, which corresponds to the green line in Fig.\ \ref{Fig:thickness_phonon}. The resulting distributions in real space are shown in Fig.\ \ref{Fig:phonon_modes} for a bandwidth $\Gamma=1$ THz. From the top-view, one can see that without gear rotation there are more atoms excited at $13$ THz than at $2$ THz, which is consistent with the features found in Fig.\ \ref{Fig:thickness_phonon} (b). To further see in detail how the mode distributes in $z$-direction, we plot the average of the kinetic energy spectrum $K_i(\nu_0)$ by spatially binning in $z$-direction with $N_{\text{bin}}=7$ bins.
Note that $z=0$ is at the bottom of the crystalline SiO$_2$ slab. The results are shown in Fig.\ \ref{Fig:vertical_distribution}. The blue and red dashed lines represent the results for $2$ THz and $13$ THz before the gear rotation, respectively. For $13$ THz, there is a plateau between $z=3$ \AA{} and $z=9$ \AA{}, which means this mode resembles a bulk normal mode
as expected. As for $2$ THz, the intensity is everywhere very low compared to $13$ THz since no mode has been excited yet.

Similarly, we repeat the procedure for the case with gear rotation. As shown in Fig.\ \ref{Fig:phonon_modes} (c) and (d), the mode at $2$ THz has been strongly excited and localizes below the gear (the red region around the center), which is also consistent with the spectra in Fig.\ \ref{Fig:thickness_phonon}. However, for $13$ THz, there is also a pattern visible, which is apparently due to the gear rotation. To get more insight, we show the resulting distribution in $z$-direction in Fig.\ \ref{Fig:vertical_distribution}. The blue and red solid lines are the results during the gear rotation. For $2$ THz (the blue solid line), the mode is stronger excited for $z>10$ \AA{}. This implies that this mode behaves like a surface phonon. As for $13$ THz (the red solid line), the intensity in the middle of the slab is reduced compared to the case without gear rotation, which is also consistent with the difference between the peak heights observed in Fig.\ \ref{Fig:thickness_phonon} (b) and (c) at $13$ THz. On the surface ($z>10$ \AA{}), however, the intensity is enhanced due to the gear rotation. Therefore, this situation corresponds to a mixture of bulk phonons and surface phonons.


\section{\label{sec:Conclusion}Conclusion and outlook}
In this study, we show that viscous dissipation can emerge from pure van-der-Waals interactions between gears and substrates. The corresponding angular velocity relaxation time can be extracted by fitting the results from the MD simulation to an exponential decay. We have also studied the effect of different gear sizes and substrate materials. The results show that when the gear size is below $3$ nm, the gear motion becomes non-smooth since it is much easier to excite the gear internal motion. Concerning the substrate, amorphous SiO$_2$ gives the largest friction coefficient, crystalline SiO$_2$ has an intermediate one, and graphene shows the smallest friction coefficient. Then, we investigated the effect of different cases with varying degree of structural flexibility (different numbers of degrees of freedom) in both gear and substrate, and find that the dissipation process is composed of different channels involving gear deformations and surface vibrations, respectively. The latter are found to be more important than gear deformations. Finally, we found that the dissipation depends on the thickness of the substrate and also directly relates to the excitation of low frequency vibrational surface modes. For the example of SiO$_2$, we found a surface phonon mode at $2$ THz to be dominant.

Future studies will need to address frictional processes at even smaller scales, involving molecule gears on metal substrates \cite{Tomassone1997,Perssson1995}. In this situation, dissipation channels involving electronic degrees of freedom become very relevant, so that computationally more expensive force fields, such as reactive force-fields (like ReaxFF) \cite{Chenoweth2008}, will be required to perform MD simulations for such systems. Finally, we hope that atomistic studies combined with further developments of the necessary fabrication tools will provide the possibility to fine tune the design of solid-state gears at the nanometer scale.

\begin{acknowledgments}
We thank A.~Kutscher, J.~Heinze, S.~Kampmann, D.~Bodesheim, N.~Lorente, C.~Joachim and F.~Moresco for useful discussions and suggestions. This work has been supported by the International Max Planck Research School (IMPRS) for ``Many-Particle Systems in Structured Environments'' and also by the European Union Horizon 2020 FET Open project ``Mechanics with Molecules'' (MEMO, grant nr.\ 766864). We also acknowledge the Center for Information Services and High Performance Computing (ZIH) at TU Dresden for computational resources.
\end{acknowledgments}

\bibliography{paper}
\end{document}